\documentclass[conference]{IEEEtran}
\IEEEoverridecommandlockouts
\usepackage{authblk}
\usepackage{amsmath,amssymb,amsfonts}
\usepackage{algorithmic}
\usepackage{graphicx}
\usepackage{textcomp}
\usepackage{xcolor}
\usepackage{lettrine}

\usepackage{amssymb}
\usepackage{pifont
}
\newcommand{\cmark}{\ding{51}}%
\newcommand{\xmark}{\ding{55}}%

\DeclareMathOperator*{\argmin}{arg\,min}

\def\BibTeX{{\rm B\kern-.05em{\sc i\kern-.025em b}\kern-.08em
    T\kern-.1667em\lower.7ex\hbox{E}\kern-.125emX}}
\begin{document}


\title{X-Vision: An augmented vision tool with real-time sensing ability in tagged environments}

\author[ ]{Yongbin Sun*\thanks{* Equal Contribution}}
\author[ ]{Sai Nithin R. Kantareddy*}
\author[ ]{Rahul Bhattacharyya}
\author[ ]{Sanjay E. Sarma}
\affil[ ]{Auto-ID Labs, Department of Mechanical Engineering }
\affil[ ]{Massachusetts Institute of Technology}
\affil[ ]{Cambridge, USA}
\affil[ ]{\{yb\_sun, nithin, rahul\_b, sesarma\} at mit.edu}

\maketitle

\begin{abstract}
    We present the concept of X-Vision, an enhanced Augmented Reality (AR)-based visualization tool, with the real-time sensing capability in a tagged environment. We envision that this type of a tool will enhance the user-environment interaction and improve the productivity in factories, smart-spaces, home \& office environments, maintenance/facility rooms and operation theatres, etc. In this paper, we describe the design of this visualization system built upon combining the object's pose information estimated by the depth camera and the object's ID \& physical attributes captured by the RFID tags. We built a physical prototype of the system  demonstrating the projection of 3D holograms of the objects encoded with sensed information like water-level and temperature of common office/household objects. The paper also discusses the quality metrics used to compare the pose estimation algorithms for robust reconstruction of the object's 3D data.
\end{abstract}
\section{Introduction}

Superimposing information on to the real-world, the concept commonly known to us as Augmented reality (AR), has been rapidly evolving over the past few years due to the advances in computer vision, connectivity and mobile computing. In recent years, multiple AR-based applications have touched everyday lives of many of us: few such examples are Google translate's augmented display \cite{good2011augmented} to improve productivity, AR GPS navigation app for travel \cite{singh2013augmented}, CityViewAR tool for tourism \cite{lee2012cityviewar}, etc. 

All these applications require a method to implement a link between the physical and digital worlds. Often this link is either ID of the object or information about the physical space, for instance, an image in Google translate app or GPS location in AR navigation tool. This link can be easily established in a informationally structured environments using visual markers, 2D barcodes and RFID tags. Among the three, RFID tags have an unique leverage with the ability to wirelessly communicate within couple of meters of distance without requiring line of sight access. In addition, RFID tags can be easily attached to inventory and consumer products in large numbers at extremely low per unit costs. 
Passive RFID, in particular, has many applications in object tracking \cite{kantareddy2017towards}, automatic inventory management \cite{kantareddy2017low}, pervasive sensing \cite{bhattacharyya2010low}, etc. In a tagged environment, with RFID infrastructure installed, information of tagged object's ID and physical attributes can be wirelessly retrieved and mapped to a digital avatar. 

\begin{figure}[t]
  \centering
  \includegraphics[width=0.45\textwidth]{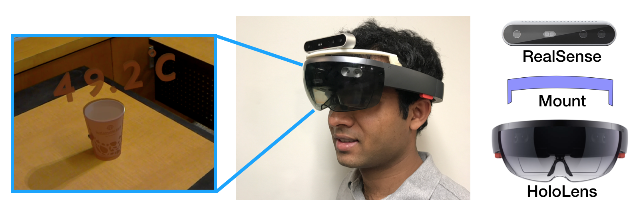}
  \caption{\textbf{Left:} A user wearing the system sees a cup with overlaid temperature information. \textbf{Right:} System components: an Intel RealSense D415 RGB-D camera is attached on a HoloLens via a custom mount.}
  \label{fig:device}
\end{figure}

In this paper, we have designed a visualization framework called \textit{X-Vision}, hoping to equip users with the ability to directly see the physical attributes of surrounding objects (Figure \ref{fig:device}). One of the goals of this framework is to demonstrate the advantages of tagged environments to enhance the user-environment interaction with real-time sensing at low cost for potential use cases in smart-spaces, home \& office environments, maintenance/facility rooms and operation theatres, etc. The rest of the paper is structured as follows: Section II discusses the relevant work in fusion of other technologies with RFID and AR; Section III provides the details of the proposed framework and the visualization system; Section IV discusses the evaluations of the test experiments, followed by the conclusions in Section V.

\begin{figure*}[h!]
  \centering
  \includegraphics[width=0.9\textwidth]{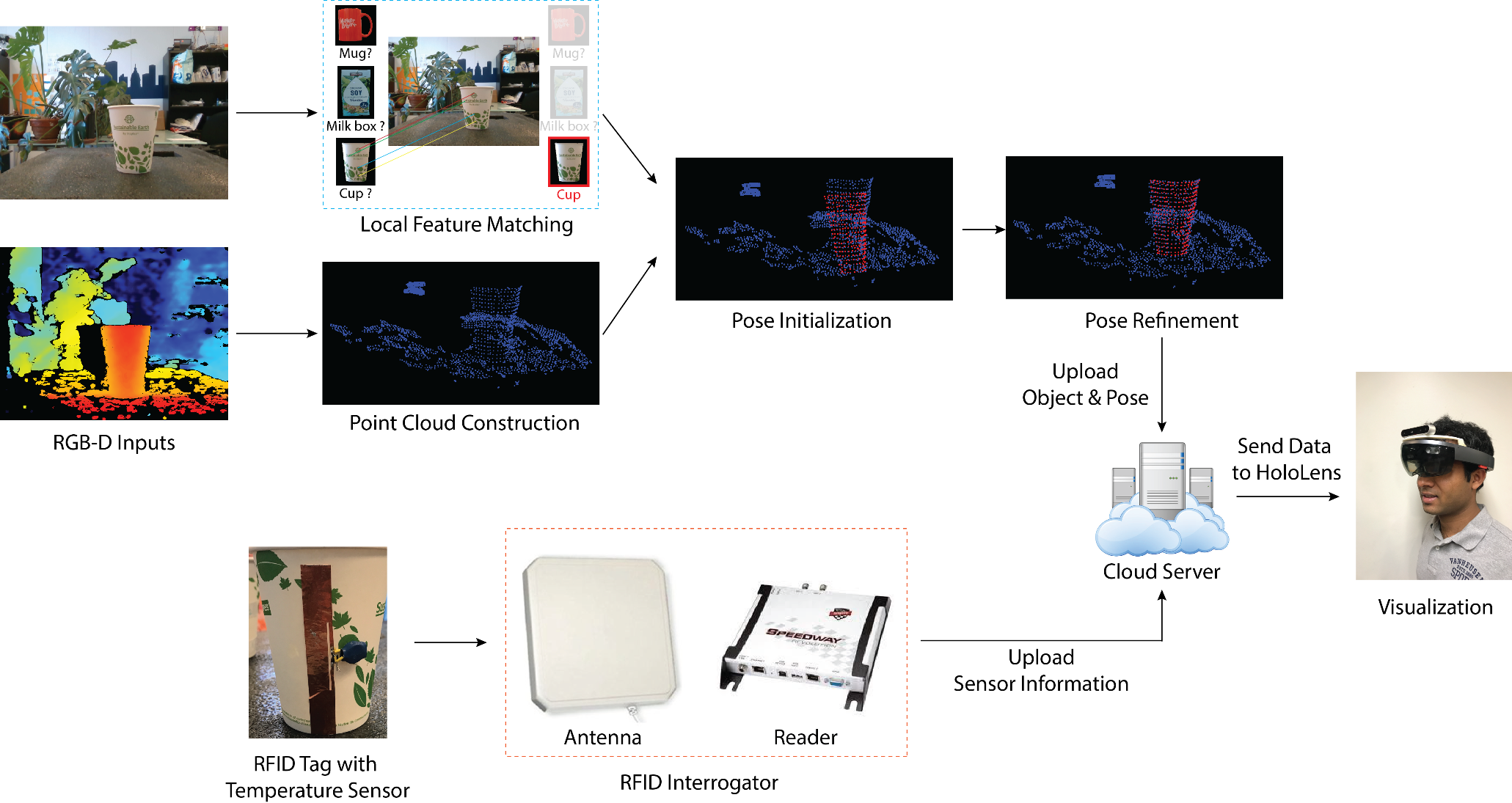}
  \caption{The pipeline of the proposed framework.}
  \label{fig:model}
\end{figure*}

\section{Related Work}

\subsection{AR-Based Smart Environment}
AR brings digital components into a person's perception of the real world. Today, advanced AR technologies facilitates the interactive bidirectional communication and control between a user and objects in the environment.
Two main branches exist for AR associated research.
In one branch, researchers attempt to design algorithms to achieve accurate object recognition and 3D pose estimation for comprehensive environment understanding.
Related work in this direction can be found in \cite{karlinsky2017fine, bay2006surf} for object recognition, and in \cite{xiang2017posecnn, wang2018dynamic, stets2017visualization} for 3D data processing. 
Research in this direction provides theoretic supports for industry products.
In the other branch, efforts have been devoted to applying existing computer vision techniques to enhance user-environment interaction experience for different purposes.
Research work on this track benefits areas, such as education \cite{garzon2017augmented}, tourism \cite{chung2015tourists} and navigation \cite{wang2014augmented}, by improving user experience.
Our work follows this trend by fusing object recognition and 3D pose estimation techniques with RFID sensing capabilities, aiming to create a smart environment.

\subsection{Emerging RFID Applications}
RFID is massively used as identification technology to support tracking in supply chain, and has so far been successfully deployed in various industries. Recently industry's focus seems shifting towards generating higher value from the existing RFID setups by tagging more \& more items and by developing new applications using tags that allow for sensing, actuation \& control \cite{ferdik2018battery} and even gaming \cite{agrawal2018tangible}. Another such exciting application with industrial benefit is fusion with emerging computer vision and AR technologies. Fusion of RFID and AR is an emerging field and there are recent studies combining these technologies for gaming and education, yet we see lot of space to explore further especially going beyond ID in RFID. One of the earlier papers \cite{rashid2006pac} studied the use of RFID to interact with physical objects in playing a smartphone-based game which enhanced the gaming experience. Another study \cite{wu2017transform} used a combination of smart bookshelves equipped with RFID tags and mixed-reality interfaces for information display in libraries. Another study \cite{ayala2015virtual} explores the use of AR with tags to teach geometry to students. These studies show a good interest in the community to explore mixed reality applications using tags for object IDs. In this paper, we use RFID for not just ID but also to wirelessly sense the environment and object's attributes to create a more intimate and comprehensive interaction between the humans and surrounding objects.

\section{System}

Our system (hardware and visualization shown in Fig. \ref{fig:device}) contains two parallel branches (shown in Figure \ref{fig:model}) to concurrently detect and sense RFID tag-sensor attached objects.
On one side, the system captures color and depth images using the depth camera for in-view target object identification and pose estimation.
On the other side, the system collects tag-data reflecting the target object's physical properties, such as temperature, using an RFID interrogator$\backslash$reader. 
Information collected from both sources are uploaded to a shared central server, where heterogeneous information is unified and delivered to the HoloLens for augmented visualization.
Details are given in the following subsections.

\subsection{Object Identification and Pose Estimation}
\label{subsec:pose}
Our system uses an Intel RealSense D415 depth camera to capture color and depth information.
It is attached to an HoloLens via a custom mount provided by \cite{garon2016real}, and faces in the same direction as the HoloLens (Figure \ref{fig:device}).
The captured images are used to identify the in-view target object and estimate its pose. 

\noindent
\textbf{Object Identification:}
Object recognition is a well-studied problem, and we adopt the local feature based method \cite{bay2006surf} in our system, since it is suitable for small-scale database. 
Generally, to identify an in-view object from a given database, the local feature based method first extracts representative local visual features for both the scene image and template object images, and then matches scene features with those of each template object. The target object in the view is identified as the template object with the highest number of matched local features. If the number of matched features of all template objects is not sufficiently large (below a predetermined threshold), then the captured view is deemed to not contain a target.
Our system follows this scheme, and uses SURF algorithm \cite{bay2006surf} to compute local features, since compared to other local feature algorithms, such as SIFT \cite{lowe1999object}, SURF is fast and good at handling images with blurring and rotation.

\noindent
\textbf{Pose Estimation:}
After identifying the in-view object, our system estimates its position and rotation, namely 3D pose, in the space, thus augmented information can be rendered properly.
We achieve this by constructing point cloud of the scene, and aligning the identified object's template point cloud with it.
Many algorithms exist for point cloud alignment, and we adapt widely-used Iterative Closest Point (ICP) algorithm \cite{besl1992method} in our system, since it usually finds a good alignment in a quick manner.
To obtain better pose estimation results, especially for non-symmetric objects (i.e. mug), a template object usually contains point clouds from multiple viewpoints. 
Yet, the performance of ICP relies on the quality of the initialization.
Our system finds a good initial pose by moving a template object's point cloud to the 3D position that is back-projected from the centroid of matched local feature coordinates in the scene image. 
The coordinates of correctly matched local feature are the 2D projections of target object surface points, thus back-projecting their centroid should return a 3D point close to target object surface points.
After initializing the pose of template point cloud, our system refines its pose using ICP.
Finally, the estimated pose can be represented as a $4\times4$ matrix, $\textbf{M}_{pose} = \textbf{M}_{ini} \textbf{M}_{icp}$, where $\textbf{M}_{ini}$ is the transformation matrix for pose initialization, and $\textbf{M}_{icp}$ is the transformation matrix for pose refinement using ICP.
All the transformation matrix are in the format of 
\begin{equation*}
 \textbf{M} = 
 \begin{bmatrix}
 \textbf{R} \quad \textbf{t} \\
 \textbf{0} \quad 1\\
 \end{bmatrix}
\end{equation*}
, where $\textbf{R}$ is a $3\times3$ matrix representing rotation, and $\textbf{t}$ is a $3\times1$ vector representing translation. Related details are illustrated in \cite{gentle2007matrix}.

\subsection{RFID Sensing}
\begin{figure}[h!]
  \centering
  \includegraphics[width=0.45\textwidth]{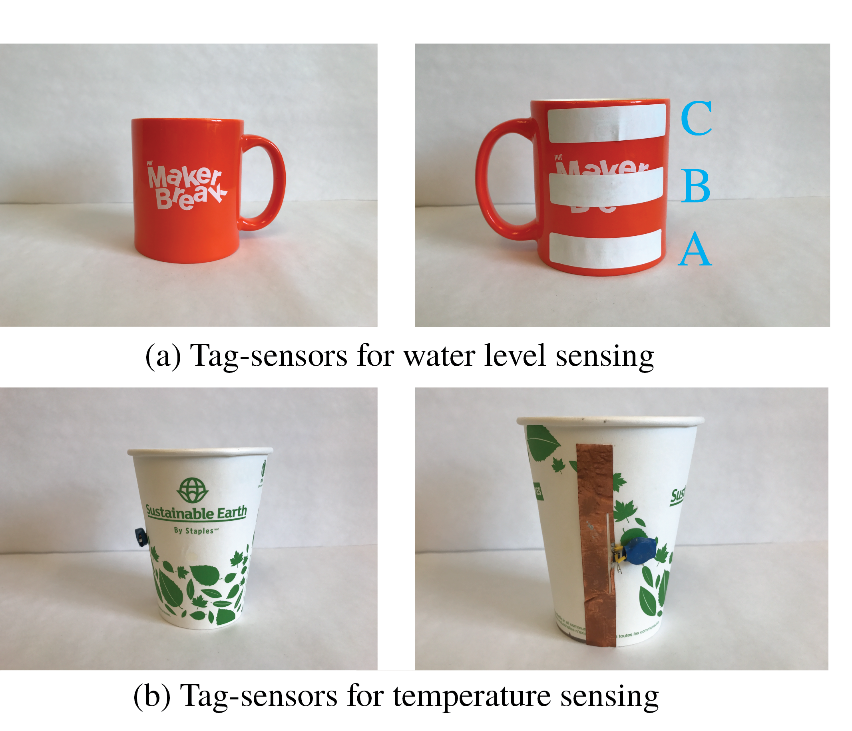}
  \vspace{-1em}
  \caption{Tag-sensors for sensing object properties: \textbf{Top row:} Water level sensing using paper tags; \textbf{Bottom row:} Temperature sensing using custom tag with EM4325.}
  \label{fig:target_object}
  \vspace{-1.8em}
\end{figure}

An office space already equipped with the RFID infrastructure is used as the tagged-environment for the experiments in this study. The space is set up using the Impinj Speedway Revolution RFID reader, connected to multiple circularly polarized Laird Antennas with gain of 8.5 dB. The reader system is broadcasting at the FCC maximum of 36 dBm EIRP.  For the tag-sensors, we make use of the Smartrac's paper RFID tags with Monza 5 IC  as the backscattered-signal based water level sensors and custom designed tags with EM 4325 IC as the temperature sensors. We use the Low Level Reader Protocol (LLRP) implemented over Sllurp (Python library) to interface with RFID readers and collect the tag-data. 

Purely-passive or semi-passive tags can be designed to sense multiple physical attributes and environmental conditions. One approach is based on tag-antenna's response to changed environments as a result of sensing event. Change in signal power or response frequency of the RFID tag due to this antenna's impedance shift can be attributed to sensing events like temperature rise \cite{vaz2010full}, presence of gas concentration \cite{occhiuzzi2011rfid}, soil moisture \cite{hasan2015towards} etc. Another approach is to use IC's on-board sensors or external sensors interfaced with GPIOs \cite{de2014battery}. In this study, we use both the antenna's impedance shift approach to detect water-level and the IC's on-board temperature sensor to detect the real-time temperature in a coffee cup.

\noindent
\textbf{Water Level Sensing:}
Water-level sensor works on the concept of relating the detuning of the tag's antenna in the presence of water in the neighborhood of the tag. In this study, we used tags as the water-level sensors on common household/office objects such as coffee cup made of paper, ceramic mug and plastic bottle. In an empty state, the background dielectric for the tags is air, therefore, the backscattered signal strength from the tags is at the maximum. In the state where the mug contains water, the antenna is significantly detuned due to the change in background dielectric, as a result the tag becomes unresponsive. However, when the mug is emptied the tag can be read again indicating empty cup. We build on this concept to detect discrete levels of water in the container by using three tags to define states as empty, low, mid, and high (illustrated in Table \ref{table:water_level}). Fig. \ref{fig:target_object}(a) shows the level sensor labels implemented on a standard ceramic coffee mug.

\begin{table}[h]
\vspace{-1em}
\caption{Water Level Indication}
\begin{center}
\begin{tabular}{|c|c|c|c|}
\hline
\textbf{Status} & \textbf{A} & \textbf{B} & \textbf{C}\\
\hline
 Empty & \xmark & \cmark & \cmark  \\
\hline
 Middle & \xmark & \xmark & \cmark \\
\hline
 Full & \xmark & \xmark  & \xmark \\
\hline
\end{tabular}
\label{table:water_level}
\end{center}
\vspace{-1em}
\end{table}

\noindent
\textbf{Temperature Sensing:}
Temperature sensor is implemented by using EM Microelectronics's EM 4325 with on-board temperature as the RFID IC. Fig. \ref{fig:target_object}(b) shows a T-match antenna with EM 4325 IC and a button-cell battery implemented as a temperature sensor on a standard coffee cup. Temperature measurements from this IC can be made in both passive as well as semi-passive mode. In the passive mode, the tag has to be in range of a reader antenna. In the semi-passive mode, the battery or external energy source keeps the IC on. The IC's temperature measurement is triggered by writing any random information into the specific word of the user memory bank (Memory bank:3 and Wordptr:256). The IC updates this word with the measured temperature from the on-board temperature sensor. By reading the word again current temperature can be known. We have implemented this code using the Sllrup library. Real-time temperature sensing is possible using this IC within $-64^o$ to $64^o$ Celsius.

\subsection{Augmented Visualization}
After obtaining the target object's identity, pose and physical properties, the system superimposes augmented information (i.e. CAD model) onto the object.
Since the object's 3D pose is estimated in the depth camera coordinate system, a series of transformations are required to obtain the 3D pose in the world coordinate system, which is required in the HoloLens rendering system.
Our system computes the transformation using:
\begin{equation*}
 \textbf{M}_{pose}^{world} = \textbf{T}_{HoloLens}^{world}
 \textbf{T}_{dep\_cam}^{HoloLens}
 \textbf{M}_{pose}^{dep\_cam}
\end{equation*}
, where $\textbf{M}_{pose}^{dep\_cam}$ and $\textbf{M}_{pose}^{world}$ are the 3D poses in depth camera and world coordinate system, respectively, $\textbf{T}_{dep\_cam}^{HoloLens}$ maps the pose from the depth camera coordinate system to the HoloLens coordinate system, and $\textbf{T}_{HoloLens}^{world}$ maps the pose from the HoloLens coordinate system to the world coordinate system. 
All the transformation matrices are in the same format as those described for pose estimation. 

\section{Evaluations}

\subsection{Sensing Results Visualization}
\begin{figure}[t!]
  \centering
  \includegraphics[width=0.45\textwidth]{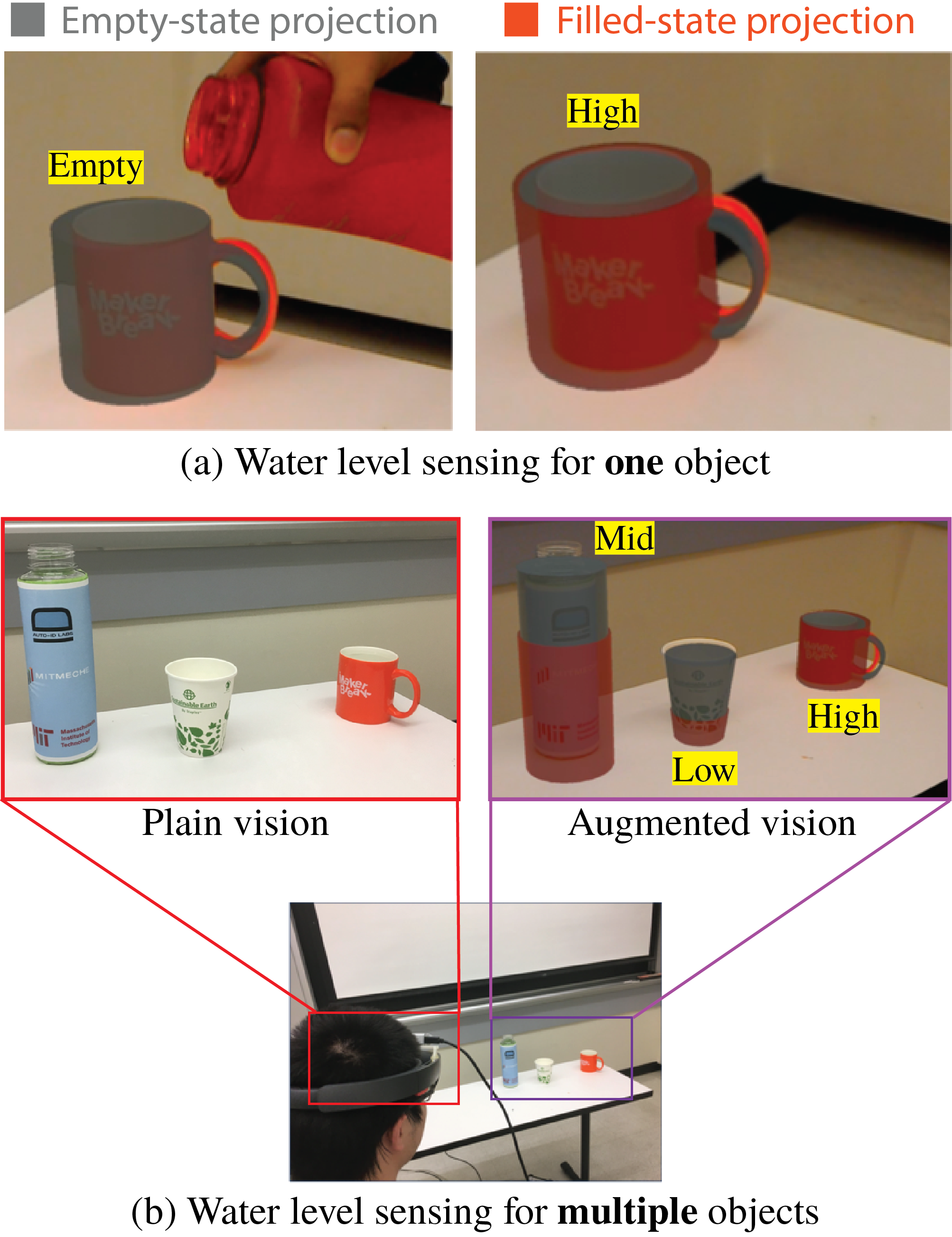}
  \vspace{-1em}
  \caption{Water level sensing results: (a) shows the HoloLens rendered results before and after water is added into a tagged mug; (b) shows multiple objects with different detected water levels.}
  \label{fig:mug_2}
  \vspace{-0.3em}
\end{figure}

\begin{figure}[t!]
  \centering
  \includegraphics[width=0.45\textwidth]{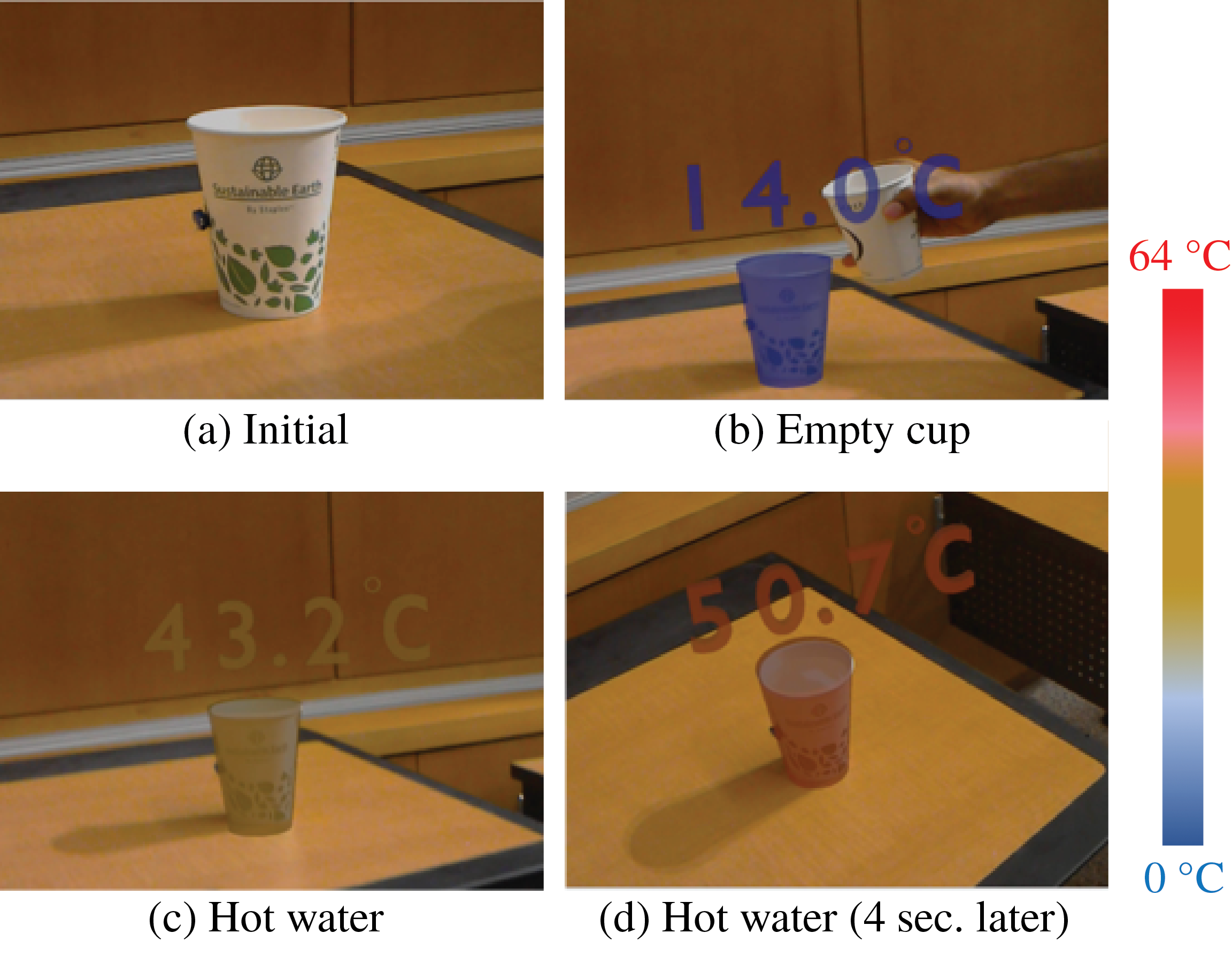}
  \vspace{-1em}
  \caption{A sequence of temperature sensing results after hot water is added. The temperature change after adding hot water results from the temperature sensor latency.}
  \label{fig:temp_2}
  \vspace{-1.9em}
\end{figure}

We first test our system's performance on both water level sensing and temperature sensing.
A user's augmented views are recorded and shown to demonstrate the effectiveness of the proposed system.

We present water level sensing results for both single object and multiple objects cases (Figure \ref{fig:mug_2}).
The system projects 3D CAD models of identified objects into the scene according to their estimated poses.
The color of projected 3D models is changed at different heights to reflect different water levels.
As can be observed, our system properly aligns 3D models to corresponding target objects, even for non-symmetric shapes (i.e. mug).
The detected discrete water levels (empty, low, mid, high) also matches the actual water level in our experiments.

Temperature sensing results are shown in Figure \ref{fig:temp_2}.
These results are selected from a recorded video clip containing the whole temperature changing process after hot water is poured into a tagged cup.
In the beginning, the tag-sensor reports room temperature for the empty cup.
After hot water is added, the tag-sensor reports water temperature (and intermediate temperatures). 
Temperatures are rendered using the color code shown on the right of Figure \ref{fig:temp_2}.
Our system shows visually appealing results.

\subsection{Pose Estimation Evaluation}
We evaluate the pose estimation pipeline used in the system by considering recognition accuracy, pose estimation quality and running time. 
The Fast Point Feature Histograms (FPFH) algorithm \cite{rusu2009fast} with and without using local features are implemented as competing methods. 
The original FPFH algorithm was designed for point cloud alignment.
As a baseline method in this experiment, it identifies the in-view object and estimates its pose by aligning each template point cloud to scene point cloud, and the object with the best alignment is selected. 
``Local Feature + FPFH" works in the same way as our pipeline by first identifying the target object using local features and then estimating its pose using FPFH.

\begin{figure}[t!]
  \centering
  \includegraphics[width=0.15\textwidth]{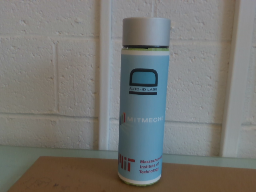}
  \includegraphics[width=0.15\textwidth]{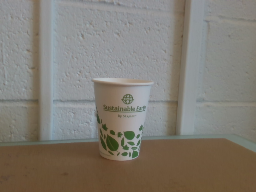}
  \includegraphics[width=0.15\textwidth]{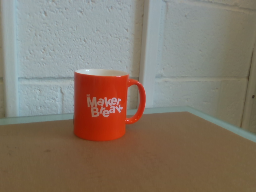}
  \caption{Test objects.}
  \label{fig:test_obj}
  \vspace{-1.8em}
\end{figure}

Three different objects (a water bottle, a coffee cup and a mug) are tested. 
For each object, five images from different angles at a range between 0.3-0.5 meter are collected for evaluation (as show in Figure \ref{fig:test_obj}).
Within this range, texture details of objects can be captured and their point clouds can be constructed with less noise.

First, the recognition accuracy of each method is evaluated and reported in Table \ref{table:acc}. As can be noted, compared to the FPFH baseline method, local visual features enhance the recognition accuracy for all cases. This performance boosting results from the rotational invariance of the detected SURF features. Identifying in-view objects is important, thus correct template point clouds can be used for pose estimation in the following step.

\vspace{-1em}
\begin{table}[h]
\caption{Recognition Accuracy}
\begin{center}
\begin{tabular}{|c|c|c|c|c|}
\hline
 & \textbf{Bottle} & \textbf{Cup} & \textbf{Mug} & \textbf{Avg.}\\
\hline
 FPFH & 1/5 & 3/5  & 3/5   & 7/15 \\
\hline
 Local Feature + ICP & \textbf{5/5} & \textbf{5/5} & \textbf{5/5} & \textbf{15/15} \\
\hline
 Local Feature + FPFH & \textbf{5/5} & \textbf{5/5} & \textbf{5/5} & \textbf{15/15} \\
\hline
\end{tabular}
\label{table:acc}
\end{center}
\end{table}

\begin{figure}[t!]
  \centering
  \includegraphics[width=0.47\textwidth]{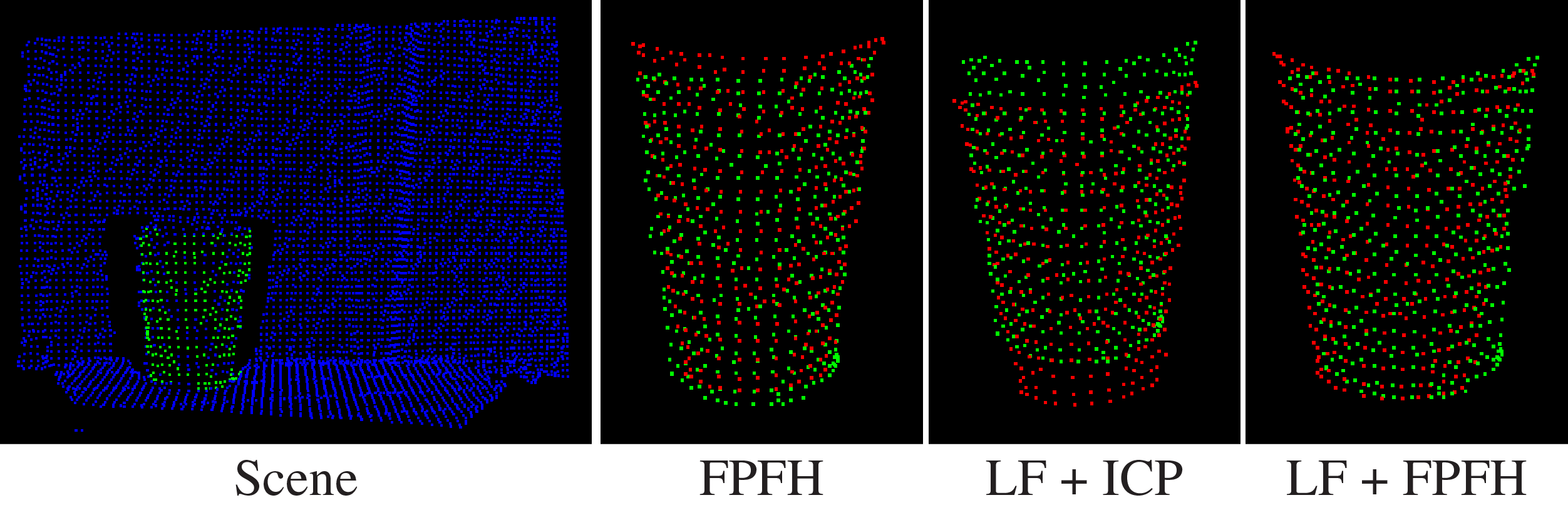}
  \caption{Point cloud pose estimation examples of different methods. Scene points are marked as blue, target object points are marked as green, and transformed template object points are marked as red.}
  \label{fig:exp2}
\end{figure}

Second, the pose estimation quality is evaluated using point-to-point residual error, which is defined as:
\begin{equation*}
\mathcal{E}= \frac{1}{n}\sum_{i=1}^n \|\boldsymbol{t}_i - \boldsymbol{p}_i\|_2 
\end{equation*}
, where $\boldsymbol{t}_i$ is the $i^{th}$ point in the target object point cloud (green points in Figure \ref{fig:exp2}), and $\boldsymbol{p}_i$ is the closest point in the transformed template object point cloud $\{\boldsymbol{p}\}$ (red points in Figure \ref{fig:exp2}) to $\boldsymbol{t}_i$, such that $\boldsymbol{p}_i = \argmin_{\boldsymbol{p}} \|\boldsymbol{t}_i - \boldsymbol{p}\|_2$.
Results are reported in Table \ref{table:error}, where the residual error is averaged across all correctly identified target objects. 
Point clouds of target objects in the scene are manually labeled (i.e. green points in Figure \ref{fig:exp2})
Due to good pose initialization from local visual feature matching, two-phase pipelines achieves lower point-to-point residual error. 
``Local Feature + FPFH" performs slightly better than ``Local Feature + ICP", since ICP is prone to get trapped into local minima.
Examples of successfully aligned point clouds are shown in Figure \ref{fig:exp2}. 

Third, we compare the running time of different methods, and report the average time for each testing object in Table \ref{table:runtime}.
Despite ``Local Feature + ICP" shows a little higher point-to-point residual error than ``Local Feature + FPFH", it runs significantly faster and is suitable for real-time performance.

\vspace{-1em}
\begin{table}[h]
\caption{Point-to-Point Residual Error}
\begin{center}
\begin{tabular}{|c|c|c|c|c|}
\hline
 & \textbf{Bottle} & \textbf{Cup} & \textbf{Mug} & \textbf{Avg.}\\
\hline
 FPFH & 0.0069 & 0.0059  & 0.0094   & 0.0075 \\
\hline
 Local Feature + ICP & 0.0088 & 0.0074 & \textbf{0.0070} & 0.0077 \\
\hline
 Local Feature + FPFH & \textbf{0.0057} & \textbf{0.0055} & 0.0087 & \textbf{0.0066} \\
\hline
\end{tabular}
\label{table:error}
\end{center}
\end{table}

\vspace{-1em}
\begin{table}[h]
\caption{Pose Estimation Time (\textit{sec.})}
\begin{center}
\begin{tabular}{|c|c|c|c|c|}
\hline
 & \textbf{Bottle} & \textbf{Cup} & \textbf{Mug} & \textbf{Avg.}\\
\hline
 FPFH & 4.603  & 4.502  & 4.590   & 4.565 \\
\hline
 Local Feature + ICP & \textbf{0.055} & \textbf{0.015} & \textbf{0.074} & \textbf{0.048} \\
\hline
 Local Feature + FPFH & 2.719 & 0.493 & 1.329 & 1.514 \\
\hline
\end{tabular}
\label{table:runtime}
\end{center}
\end{table}
\vspace{-1.2em}

\subsection{Working Range Testing}
Object recognition accuracy of our system is affected by the camera-object separation. The minimum distance recommended by the depth camera manufacturer is 30 cm. As the separation increases, the quality of the depth data deteriorates ，and beyond 1 m, texture details of target objects are hard to capture. Similarly, RFID tag-reader communication is affected by the tag-reader separation. If the separation is too large, the power reaching the tag is too low to power the IC and backscatter the signal to the reader. We define a score called normalized RSSI for generalized comparison between different material-range-signal strength experiments. Score of 1 denotes a good backscattered signal strength of -20 dBm at the reader and a score of 0 means signal strength is below the sensitivity of the reader (-80 dBm).

\begin{figure}[h!]
  \centering
  \includegraphics[width=0.47\textwidth]{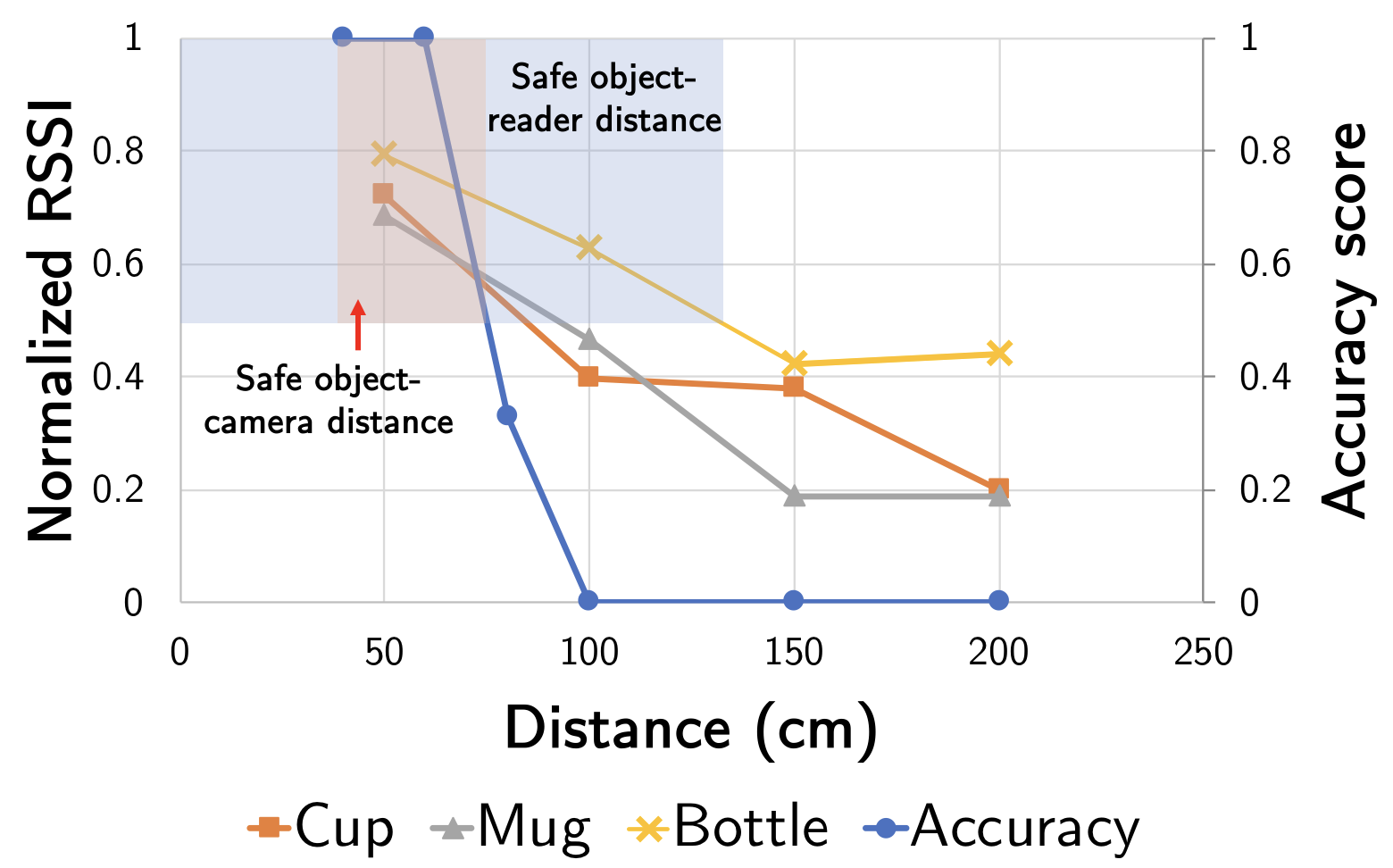}
  \caption{Plot showing normalized detection parameters (normalized RSSI in case of RFID and projection accuracy in case of augmented vision) from 1: good to 0: poor; 0.5 is chosen as the safe limit for good quality detection and rendering. Shaded regions show safe distances for object \& RFID-reader and object \& depth-camera to achieve good quality results}
  \label{fig:comp}
\end{figure}

Recognition accuracy and normalized RSSI scores are obtained for different objects in this study by varying the camera-object and reader-object separation distances (see Fig.\ref{fig:comp}). From our observations, to achieve a reliable sensing and good quality visualization, we set an acceptable score of 0.5-1 for both the metrics. 
We propose a 40-75 cm working range between the camera \& target object, and less than 100-150 cm working range between the tagged objects \& readers for good quality and reliable visualization. One of our ultimate goals is to package the camera and reader on to the head mount so that a need for separate RFID infrastructure is eliminated. Therefore, this data shows that the RFID range is suitable for this type of application and the human-object distance is limited by the camera.

\section{Conclusion}
We present the working of an enhanced augmented-vision system named X-Vision which superimposes physical objects with 3D holograms encoded with sensing information captured from the tag-sensors attached to everyday objects.
Two testing cases, water level and temperature sensing, are demonstrated in this paper.
We also observe that the distance between the depth camera and RFID reader with the objects is critical for system performance. 
We propose a 40-75 cm range between the camera \& target object, and less than 100-150 cm between the tagged objects \& readers for good quality and reliable visualization.

\bibliographystyle{unsrt}
\bibliography{reference}

\begin{thebibliography}{10}

\bibitem{good2011augmented}
Otavio Good.
\newblock Augmented reality language translation system and method, April~21
  2011.
\newblock US Patent App. 12/907,672.

\bibitem{singh2013augmented}
Mona Singh and Munindar~P Singh.
\newblock Augmented reality interfaces.
\newblock {\em IEEE Internet Computing}, 17(6):66--70, 2013.

\bibitem{lee2012cityviewar}
Gun~A Lee, Andreas D{\"u}nser, Seungwon Kim, and Mark Billinghurst.
\newblock Cityviewar: A mobile outdoor ar application for city visualization.
\newblock In {\em Mixed and Augmented Reality (ISMAR-AMH), 2012 IEEE
  International Symposium on}, pages 57--64. IEEE, 2012.

\bibitem{kantareddy2017towards}
SNR Kantareddy, R~Bhattacharyya, and S~Sarma.
\newblock Towards low-cost object tracking: Embedded rfid in golf balls using
  3d printed masks.
\newblock In {\em RFID (RFID), 2017 IEEE International Conference on}, pages
  137--143. IEEE, 2017.

\bibitem{kantareddy2017low}
SNR Kantareddy, R~Bhattacharyya, and SE~Sarma.
\newblock Low-cost, automated inventory control of sharps in operating theaters
  using passive rfid tag-sensors.
\newblock In {\em RFID Technology \& Application (RFID-TA), 2017 IEEE
  International Conference on}, pages 16--21. IEEE, 2017.

\bibitem{bhattacharyya2010low}
Rahul Bhattacharyya, Christian Floerkemeier, and Sanjay Sarma.
\newblock Low-cost, ubiquitous rfid-tag-antenna-based sensing.
\newblock {\em Proceedings of the IEEE}, 98(9):1593--1600, 2010.

\bibitem{karlinsky2017fine}
Leonid Karlinsky, Joseph Shtok, Yochay Tzur, and Asaf Tzadok.
\newblock Fine-grained recognition of thousands of object categories with
  single-example training.
\newblock In {\em Proceedings of the IEEE Conference on Computer Vision and
  Pattern Recognition}, pages 4113--4122, 2017.

\bibitem{bay2006surf}
Herbert Bay, Tinne Tuytelaars, and Luc Van~Gool.
\newblock Surf: Speeded up robust features.
\newblock In {\em European conference on computer vision}, pages 404--417.
  Springer, 2006.

\bibitem{xiang2017posecnn}
Yu~Xiang, Tanner Schmidt, Venkatraman Narayanan, and Dieter Fox.
\newblock Posecnn: A convolutional neural network for 6d object pose estimation
  in cluttered scenes.
\newblock {\em arXiv preprint arXiv:1711.00199}, 2017.

\bibitem{wang2018dynamic}
Yue Wang, Yongbin Sun, Ziwei Liu, Sanjay~E Sarma, Michael~M Bronstein, and
  Justin~M Solomon.
\newblock Dynamic graph cnn for learning on point clouds.
\newblock {\em arXiv preprint arXiv:1801.07829}, 2018.

\bibitem{stets2017visualization}
Jonathan~Dyssel Stets, Yongbin Sun, Wiley Corning, and Scott~W Greenwald.
\newblock Visualization and labeling of point clouds in virtual reality.
\newblock In {\em SIGGRAPH Asia 2017 Posters}, page~31. ACM, 2017.

\bibitem{garzon2017augmented}
Juan Garz{\'o}n, Juan Pav{\'o}n, and Silvia Baldiris.
\newblock Augmented reality applications for education: Five directions for
  future research.
\newblock In {\em International Conference on Augmented Reality, Virtual
  Reality and Computer Graphics}, pages 402--414. Springer, 2017.

\bibitem{chung2015tourists}
Namho Chung, Heejeong Han, and Youhee Joun.
\newblock Tourists’ intention to visit a destination: The role of augmented
  reality (ar) application for a heritage site.
\newblock {\em Computers in Human Behavior}, 50:588--599, 2015.

\bibitem{wang2014augmented}
Junchen Wang, Hideyuki Suenaga, Kazuto Hoshi, Liangjing Yang, Etsuko Kobayashi,
  Ichiro Sakuma, and Hongen Liao.
\newblock Augmented reality navigation with automatic marker-free image
  registration using 3-d image overlay for dental surgery.
\newblock {\em IEEE transactions on biomedical engineering}, 61(4):1295--1304,
  2014.

\bibitem{ferdik2018battery}
Manuel Ferdik, Georg Saxl, and Thomas Ussmueller.
\newblock Battery-less uhf rfid controlled transistor switch for internet of
  things applications—a feasibility study.
\newblock In {\em Wireless Sensors and Sensor Networks (WiSNet), 2018 IEEE
  Topical Conference on}, pages 96--98. IEEE, 2018.

\bibitem{agrawal2018tangible}
Ankur Agrawal, Glen~J Anderson, Meng Shi, and Rebecca Chierichetti.
\newblock Tangible play surface using passive rfid sensor array.
\newblock In {\em Extended Abstracts of the 2018 CHI Conference on Human
  Factors in Computing Systems}, page D101. ACM, 2018.

\bibitem{rashid2006pac}
Omer Rashid, Will Bamford, Paul Coulton, Reuben Edwards, and Jurgen Scheible.
\newblock Pac-lan: mixed-reality gaming with rfid-enabled mobile phones.
\newblock {\em Computers in Entertainment (CIE)}, 4(4):4, 2006.

\bibitem{wu2017transform}
Ko-Chiu Wu, Chun-Ching Chen, Tzu-Heng Chiu, and I-Jen Chiang.
\newblock Transform children's library into a mixed-reality learning
  environment: Using smartwatch navigation and information visualization
  interfaces.
\newblock In {\em Pacific Neighborhood Consortium Annual Conference and Joint
  Meetings (PNC), 2017}, pages 1--8. IEEE, 2017.

\bibitem{ayala2015virtual}
Andr{\'e}s Ayala, Graciela Guerrero, Juan Mateu, Laura Casades, and Xavier
  Alam{\'a}n.
\newblock Virtual touch flystick and primbox: two case studies of mixed reality
  for teaching geometry.
\newblock In {\em International Conference on Ubiquitous Computing and Ambient
  Intelligence}, pages 309--320. Springer, 2015.

\bibitem{garon2016real}
Mathieu Garon, Pierre-Olivier Boulet, Jean-Philippe Doironz, Luc Beaulieu, and
  Jean-Fran{\c{c}}ois Lalonde.
\newblock Real-time high resolution 3d data on the hololens.
\newblock In {\em Mixed and Augmented Reality (ISMAR-Adjunct), 2016 IEEE
  International Symposium on}, pages 189--191. IEEE, 2016.

\bibitem{lowe1999object}
David~G Lowe.
\newblock Object recognition from local scale-invariant features.
\newblock In {\em Computer vision, 1999. The proceedings of the seventh IEEE
  international conference on}, volume~2, pages 1150--1157. Ieee, 1999.

\bibitem{besl1992method}
Paul~J Besl and Neil~D McKay.
\newblock Method for registration of 3-d shapes.
\newblock In {\em Sensor Fusion IV: Control Paradigms and Data Structures},
  volume 1611, pages 586--607. International Society for Optics and Photonics,
  1992.

\bibitem{gentle2007matrix}
James~E Gentle.
\newblock {\em Matrix transformations and factorizations}.
\newblock Springer, 2007.

\bibitem{vaz2010full}
Alexander Vaz, Aritz Ubarretxena, Ibon Zalbide, Daniel Pardo, H{\'e}ctor Solar,
  Andr{\'e}s Garcia-Alonso, and Roc Berenguer.
\newblock Full passive uhf tag with a temperature sensor suitable for human
  body temperature monitoring.
\newblock {\em IEEE Transactions on Circuits and Systems II: Express Briefs},
  57(2):95--99, 2010.

\bibitem{occhiuzzi2011rfid}
Cecilia Occhiuzzi, Amin Rida, Gaetano Marrocco, and Manos Tentzeris.
\newblock Rfid passive gas sensor integrating carbon nanotubes.
\newblock {\em IEEE Transactions on Microwave Theory and Techniques},
  59(10):2674--2684, 2011.

\bibitem{hasan2015towards}
Azhar Hasan, Rahul Bhattacharyya, and Sanjay Sarma.
\newblock Towards pervasive soil moisture sensing using rfid tag antenna-based
  sensors.
\newblock In {\em RFID Technology and Applications (RFID-TA), 2015 IEEE
  International Conference on}, pages 165--170. IEEE, 2015.

\bibitem{de2014battery}
Danilo De~Donno, Luca Catarinucci, and Luciano Tarricone.
\newblock A battery-assisted sensor-enhanced rfid tag enabling heterogeneous
  wireless sensor networks.
\newblock {\em IEEE Sensors Journal}, 14(4):1048--1055, 2014.

\bibitem{rusu2009fast}
Radu~Bogdan Rusu, Nico Blodow, and Michael Beetz.
\newblock Fast point feature histograms (fpfh) for 3d registration.
\newblock In {\em Robotics and Automation, 2009. ICRA'09. IEEE International
  Conference on}, pages 3212--3217. IEEE, 2009.

\end{thebibliography}

\end{document}